\documentclass[aps.prb,twocolumn,showpacs,preprintnumbers,amsmath,amssymb]{revtex4}
\usepackage{mathrsfs}
\usepackage{graphicx}
\usepackage{amsmath}

\begin{document}

\title{Vortex quantum tunnelling versus thermal activation
in ultrathin superconducting nanoislands}
\author{W. V. Pogosov$^{1,2}$ and V. R. Misko$^{1}$}
\affiliation{$^{1}$Departement Fysica, Universiteit Antwerpen,
Groenenborgerlaan 171, B-2020 Antwerpen, Belgium}
\affiliation{$^{2}$Institute for Theoretical and Applied
Electrodynamics, Russian Academy of Sciences, Izhorskaya 13,
125412, Moscow, Russia}

\begin{abstract}
We consider two possible mechanisms for single-vortex fluctuative
entry/exit through the surface barrier in ultrathin
superconducting disk-shaped nanoislands made of Pb and consisting
of just few monoatomic layers, which can be fabricated using
modern techniques. We estimate tunnelling probabilities and
establish criteria for the crossover between these two mechanisms
depending on magnetic field and system sizes. For the case of
vortex entry, quantum tunnelling dominates on the major part of
the temperature/flux phase diagram. For the case of vortex exit,
thermal activation turns out to be more probable. This nontrivial
result is due to the subtle balance between the barrier height and
width, which determine rates of the thermal activation and quantum
tunnelling, respectively.
\end{abstract}

\author{}
\maketitle
\date{\today }

Recent progress in miniaturization techniques has made it possible
to create ultrathin quasi-two-dimensional superconducting
structures, which consist of just few monoatomic layers, and even
one-layer films. In contrast to previously studied mesoscopic
superconductors~\cite{Geim}, both thermal and quantum fluctuations
can be expected to be important in such extremely thin
structures~\cite{Varlamov}. In addition, strong confinement of
condensate in lateral dimensions leads to a rather rich variety of
possible behaviors. In Ref. \cite{Japanese}, vortices in
superconducting nanoislands of Pb with thickness of few nanometers
have been studied experimentally. It was observed that hysteresis
effects are significantly weaker than expected from the theory. In
another experiment \cite{Dima}, similar islands have been
explored. The experiment has revealed a full suppression of the
surface barrier \cite{BL} for the \textit{single}-vortex
entry/exit. These results (see also Refs. \cite{Bartolf,Bul})
suggest that some kind of fluctuations of the superconducting
state might be responsible for the observed features. Indeed, it
was show in Ref. \cite{Pogosov} that the surface barrier height,
under the conditions of the experiment \cite{Dima}, is low enough
to be suppressed by thermal fluctuations.

The aim of the present paper is to consider another possible
mechanism of surface barrier suppression in ultrathin
superconducting nanoislands, which is due to quantum fluctuations,
as well as to establish a dominant mechanism, i.e., thermal
activation (TA) or quantum tunnelling (QT). We consider a
disk-shaped superconducting island consisting of just few
monoatomic layers with radius much larger than the
zero-temperature coherence length. The magnetic field is applied
perpendicular to the island surface. Presence of the boundaries
results in the surface barrier which prevents vortex entry or
exit. In the absence of fluctuations, the barrier for vortex entry
disappears, when the magnetic field reaches some critical value.
The barrier for the vortex exit is suppressed at a lower magnetic
field. The difference between these two fields thus reflects the
significance of hysteresis effects. In the London model, which
remains applicable down to low temperatures, these two critical
fields can be calculated by using the approach developed by Buzdin
\cite{Buzdin}. Note that macroscopic QT of Abrikosov vortices has
been studied both experimentally and theoretically for the case of
long current-carrying strip in zero applied field, where vortex
penetration is due to the current \cite{Glazman,Glazman1,Kogan},
as well as for bulk superconductors, where energy barriers appear
due to the pinning of vortices by spatial inhomogeneities
\cite{RevModPhys} or due to intrinsic pinning induced by the
layered structure \cite{Ivlev} (for the QT of Josephson vortex,
see recent papers \cite{Goldobin}). In the present paper, we apply
a combination of known approaches to the particular case of the
ultrathin mesoscopic disks, which became experimentally testable
quite recently. Apart from the interest from the viewpoint of a
possible experimental examination, it is also attracting to see
how strong confinement in lateral dimensions affects an interplay
between quantum and thermal scenarios, since it is known that
probabilities of these processes are controlled by two different
parameters, which are barrier width and barrier
height~\cite{Larkin,Glazman,RevModPhys}, these parameters being
strongly dependent on system sizes and geometry.

In order to estimate the probability of macroscopic QT of
Abrikosov vortex to/from the island, we use an approach developed
by Caldeira and Leggett \cite{Leggett}, which allows one to take
into account energy dissipation. For a vortex with mass $m_{v}$ in
a potential $V(r)$ (to be found within the London model), the
Euclidean action is given by \cite{Leggett}
\begin{equation}
S_{E}=\int dt\left[ \frac{m_{v}}{2}(\overset{.}{r})+V(r)\right] +
\frac{\eta }{4\pi }\int dt\int dt^{\prime }\frac{[r(t)-r(t^{\prime })]^{2}}{%
(t-t^{\prime })^{2}}.\label{action}
\end{equation}%
The last term in the right-hand side (r.h.s.) of Eq.
(\ref{action}), which is nonlocal, takes into account the energy
dissipation. In the reciprocal space, we obtain
\begin{equation}
S_{E}=\int \frac{d\omega }{2\pi }\left\{ \frac{m_{v}}{2}\left[ 1+\frac{\eta }{%
|\omega |m_{v}}\right] \omega ^{2}|u(\omega )|^{2}+V(u)\right\} .
\label{actionF}
\end{equation}%
It is convenient to introduce a frequency-dependent effective mass
as $m_{eff}(\omega )=m_{v}+\eta / |\omega |$. Note that it is
shown below that mass $m_{v}$ is much smaller than the
contribution from the viscosity.

In principle, one can use Eq. (\ref{actionF}) to find optimal
$u(\omega )$ exactly. However, it is known that quite accurate
results for the effective action can be obtained by using
dimensional estimates\cite{RevModPhys} (in addition, we also use
the method of Ref. \cite{Ivlev}). Following Refs.
\cite{Glazman,RevModPhys}, we find the characteristic tunnelling
frequency $\omega _{0}$ from the balance of kinetic and potential
energies as
\begin{equation}
m_{eff}(\omega _{0})l_{b}^{2}\omega _{0}^{2}=V_{b},  \label{freq}
\end{equation}%
where $V_{b}$ and $l_{b}$ are barrier height and width,
respectively. Eq. (\ref{freq}) is a quadratic equation, from which
$\omega _{0}$ can be found. In the viscous limit ($m_{v}V_{b}\ll
\eta ^{2}l_{b}^{2}$), the solution reads as
\begin{equation}
\omega _{0}\approx \frac{V_{b}}{\eta l_{b}^{2}}\left[
1-\frac{m_{v}V_{b}}{\eta ^{2}l_{b}^{2}}\right] , \label{freqvis}
\end{equation}%
where we have kept a linear term. It will be seen below that the
dissipative regime is indeed relevant to the situation we consider
here. Using Eq. (\ref{actionF}) and (\ref{freqvis}), we find the
action as
\begin{equation}
S_{E}\approx 2\eta l_{b}^{2}\left( 1+\frac{mV_{b}}{\eta ^{2}l_{b}^{2}}%
\right) .  \label{act}
\end{equation}
Note that in the case of quadric-plus-cubic potential, which was
solved exactly in Ref. \cite{Larkin}, there appears a prefactor
$\pi/2$ instead of 2 in the dominant term of the r.h.s. of Eq.
(\ref{act}), which demonstrates a high accuracy of the method of
dimensional estimates.

The viscosity coefficient for the quasi-two-dimensional
superconductor, in the case of Ohmic dissipation, is
\begin{equation}
\eta=\frac{\Phi_{0}^{2}}{2\pi\xi(T)^{2}}\frac{1}{R_{N}},
\label{visc}
\end{equation}
where $R_{N}$ is a two-dimensional resistance, which depends on
island thickness $d$ as\cite{Glazman1} $R_{N}=R_{0}d_{0}/d$, where
$d_{0}$ is the thickness of one monolayer. According to Ref.
\cite{Glazman1}, $R_{0}$ is nearly equal to $6$ kOhm for Pb.

The mass of a vortex can be estimated using the well-known
expression by Suhl~\cite{Suhl}. It includes two contributions: (i)
due to the kinetic energy of the vortex core and (ii) due to the
electromagnetic energy. In Ref.~\cite{Chudn}, another contribution
was found, which was attributed to the shear deformations of the
crystal lattice. This additional contribution is typically of the
same order as those in Ref.~\cite{Suhl}, so that it does not
violate viscous-limit conditions. These conclusions apply to the
dirty-limit regime, which is realized for superconducting
nanostructures placed on a disordered substrate, as relevant for
the experiments~\cite{Dima,Japanese}. In contrast, in the clean
limit, one has to take into account a much larger vortex mass
arising from the quantization of the electron states in the vortex
core~\cite{Kopnin}, which may depend on the frequency of the
external drive. In the dirty limit, however, it becomes of the
order of Suhl mass~\cite{Sonin,Otterlo,Chudn}.

The probability of QT, within exponential accuracy, is $P_{QT}\sim
\exp (-S_{E}/\hbar )$, while the probability of TA is $P_{TA}\sim
\exp (-V_{b}/k_{B}T)$.

The effective action, in the dissipative limit, is mainly due to
the width of the barrier, as follows from Eq. (\ref{act}) (see
also Refs. \cite{Glazman,RevModPhys}). Thus, the probability of QT
is controlled by barrier {\it width}. At the same time, the
probability of TA is determined predominantly by the barrier { \it
height}. Below we show that the subtle balance between the two
characteristics of the barrier leads to a rather rich behavior of
the system we here study.

In the London model, the energy as a function of the vortex
position $\rho $ can be found analytically \cite{Buzdin}. For a
disk of radius $R$ much smaller than the effective penetration
depth $\lambda _{\perp}=\lambda (T)^{2}/d$, the position-dependent
part of the energy reads as \cite{Buzdin}:
\begin{equation}
U(\rho )=\frac{\Phi _{0}^{2}d}{8\pi ^{2}\lambda (T)^{2}}\left\{ \ln \frac{%
R^{2}-\rho ^{2}}{R\xi (T)}+\frac{\Phi }{\Phi _{0}}\left(\frac{\rho ^{2}}{%
R^{2}}\right)\right\} ,  \label{Buzdin}
\end{equation}%
where $\Phi$ is a magnetic flux through the disk.

%In the present paper,
We consider a superconducting island placed on a disordered
substrate\cite{Dima,Japanese}. Thus, the diffusive regime is
realized with the mean free path nearly equal to $2d$. It was
shown experimentally in Ref. \cite{Ozera} that the Ginzburg-Landau
theory remains applicable for such extremely thin films. For
$\lambda (T)$ and $\xi (T)$, we use the diffusive-limit
expressions given by
\begin{eqnarray}
\lambda (T)=0.6\lambda _{0}\sqrt{\frac{\xi _{0}/2d}{1-T/T_{c}}},
  \xi (T)=%
\sqrt{\frac{2\xi _{0}d}{1-T/T_{c}}},
  \label{lambdaksi}
\end{eqnarray}
where bulk $\lambda _{0}=40$ nm and $\xi _{0}=80$ nm for Pb. For
the dependence of $T_{c}$ on $d$, we utilize the expression
deduced from the experiments on few-monolayer Pb films
\cite{Ozera}:
\begin{equation}
T_{c}(d)=T_{c0}\left( 1-\frac{d_{c}}{d}\right) ,  \label{Tc}
\end{equation}%
where $T_{c0}=7.2$ K is the critical temperature of Pb in a bulk,
while the "critical thickness" $d_{c}$ corresponds nearly to 1.5
monolayers.

It is easy to show from Eq. (\ref{Buzdin}) that the position of
the saddle point, which separates vortex-free and one-vortex
states, is
\begin{equation}
\rho _{c}=R\left( \frac{\Phi -\Phi _{0}}{\Phi }\right) ^{1/2}.  \label{roc}
\end{equation}%
The condition for the vortex entry can be obtained by equating
$\rho _{c}$ to $R-\xi (T)$. After simple algebra, we obtain the
critical flux $\Phi _{in}$, at which barrier for vortex entry
disappears
\begin{equation}
\frac{\Phi _{in}}{\Phi _{0}}=\frac{R}{2\xi (T)}\frac{1}{1-\frac{\xi (T)}{2R}}%
.  \label{Phiin}
\end{equation}%
When magnetic flux is lower than $\Phi _{in}$, the barrier height
$U_{in}$ for the vortex entry is nonzero. It can be calculated as
the difference in energies in a saddle-point configuration and at
$\rho =R-\xi (T)$:
\begin{eqnarray}
U_{in}&=&\frac{\Phi _{0}^{2}d}{8\pi ^{2}\lambda (T)^{2}}\left\{ \ln
\frac{R}{\xi (T)\left( 2-\frac{\xi (T)}{R}\right) }-\ln \frac{\Phi
}{\Phi _{0}}\right. \nonumber \\
&+& \left.\frac{\Phi }{\Phi _{0}}\left( \frac{2\xi
(T)}{R}-\frac{\xi (T)^{2}}{R^{2}}\right) -1 \right\}.
\label{Uin}
\end{eqnarray}

The condition for vortex exit is $\rho_{c}=0$, which immediately
leads to the critical flux $\Phi_{out}= \Phi_{0}$. When $\Phi >
\Phi_{out}$, the barrier height $U_{out}$ for vortex exit is
nonzero. It can be found as a difference of energies in a
saddle-point state and at $\rho=0$:
\begin{equation}
U_{out}=\frac{\Phi _{0}^{2}d}{8\pi ^{2}\lambda (T)^{2}}
\left\{
%\left\{
\left[
\frac{\Phi}{\Phi_{0}}-1
%\right\}
\right]
-\ln
\frac{\Phi}{\Phi_{0}}\right\}. \label{Uout}
\end{equation}

Thus, it is possible to find analytically barrier heights both for
%the
vortex entry and exit. However, it is not possible to find the
barrier widths analytically, since it amounts solving
transcendental equations. Namely, the barrier width for vortex
entry is given by the difference in two values of $\rho$, the
first one being $R-\xi(T)$, while the second one corresponding to
another solution of equation $U(\rho)=U(R-\xi(T))$. Similarly, the
barrier width for vortex exit is given by a nonzero value of
$\rho$, at which $U(\rho)=U(0)$. The above two equations are then
solved numerically.

Let us focus on the case of an island with $R=120$~nm and
$d=1$~nm. These are realistic parameters which are available in
modern experiments. The disk radius in terms of $\xi(0)$ is 10,
while $d$ is nearly 3 monolayers. This leads to $T_{c}=3.6$ K. In
Fig. 1, we show typical energy profiles for vortex entry/exit. It
is quite remarkable that the width of the barrier for the entry is
generally much smaller than that for the exit. Since the
probability of QT is controlled by the \textit{square} of the
barrier width, we may expect that such an asymmetry can provoke
thermal escape of a vortex from the island to be more probable
than QT.

\begin{figure}
\includegraphics[width=9.0cm]{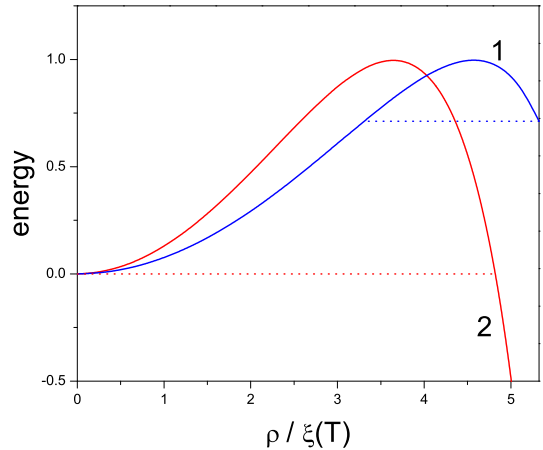}\\
\caption{
(Color online)
Typical curves for the energy of the system
as a function of the vortex position in a superconducting island
of radius 120~nm and thickness of 1~nm. Curve 1 describes the
barrier for vortex entry at $\Phi=2.1\Phi_{0}$ and $T=0.6T_{c}$.
Curve 2 corresponds to the barrier for vortex exit at
$\Phi=1.5\Phi_{0}$ and $T=0.6T_{c}$. The energies are normalized
by their values at saddle points. Dotted lines are guides for eyes used to depict barrier widths.
}
\label{fig01}
\end{figure}

Now we are going to address two issues. First, we would like to
understand if TA and/or QT is realistic for the system we
consider. Second, we want to establish, which mechanism is
dominant depending on $T$ and $\Phi$.

Our calculations show that exponents for both QT and TA turn out
to be $\sim 10-100$ almost for the whole phase diagram. We believe
that values close to ten or even several times larger are
sufficient for the vortex to overcome the barrier: Indeed, it was
shown in Refs.~\cite{Burlachkov, Koshelev, Pogosov} by evaluating
the pre-exponential factors that similar values of the exponent
result in observable rates for TA for both bulk and nanosized
superconductors. In addition, we ensure that we are indeed in the
dissipative limit, since $m_{v}V_{b}/ \eta ^{2}l_{b}^{2}$ is
several orders of magnitude smaller than 1.

\begin{figure}
\includegraphics[width=9.0cm]{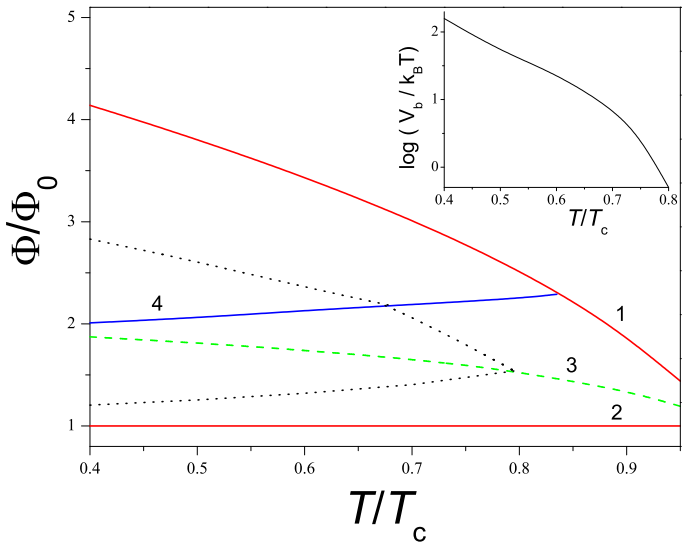}\\
\caption{ (Color online) The ``phase diagram'' ``Magnetic field
vs. Temperature'' for the island (see the text). The inset shows
the dependence of the decimal logarithm of $S_{E}/\hbar$ on
$T/T_{c}$ along curve 4.} \label{fig02}
\end{figure}

Our results are summarized in Fig. 2. Curve 1 shows $\Phi_{in}$ as
a function of $T$. Curve 2 represents $\Phi_{out}$. Dashed curve 3
gives $\Phi$, at which the energies of the state with vortex at
the disk center and at $\rho =R-\xi (T)$ are equal. For the domain
on the phase diagram, which is between curves 3 and 1, vortex
entry is energetically favorable. For the region between curves 3
and 2, vortex exit is preferable.

Curve 4 separates two domains: below this curve change of the
topological charge is more probable through thermal fluctuations,
while above this curve quantum fluctuations dominate. We denote
the corresponding temperature as $T_{0}$, while $k_{B}T_{0}=(\hbar
/2 \eta) (V_{b}/l_{b}^{2})$. For the domain between curves 4 and
1, QT of the vortex into the disk is more probable than TA. We see
that the width of this region expands with decreasing $T$, which
is reasonable. For the region confined between curves 1, 3, and 4,
TA of vortex into the disk is more probable. It is remarkable that
this scenario survives down to rather low $T$. The domain between
curves 2 and 3 corresponds to the vortex escape from the island.
In contrast to the case of vortex exit, it is fully determined by
the thermal mechanism, even at relatively low $T$. This is due to
the fact that barrier width for exit is quite large and is
controlled mostly by the disk radius, while the one for the entry
is mainly determined by $\xi(T)$ (see Fig. 1). The inset in Fig. 2
shows the dependence of $V_{b}/k_{B}T$ on $T/T_{c}$ along curve 4,
where it is equal to $S_{E}/\hbar$. It is also of interest that
curve 4 is actually oriented nearly horizontally in this ``phase
diagram''. It crosses curve 1, which corresponds to the threshold
for the vortex entry in the absence of fluctuations, at $T \approx
0.85 T_{c}$. This implies that, along curve 1, QT remains more
probable up to quite high $T$.

Finally, two dotted lines in Fig. 2 depict borders of the regions,
close to the threshold values of $\Phi$, where the exponent for
the dominant mechanism of vortex entry/exit is lower than 10, so
that a change in the system's topological charge becomes highly
probable. These lines can be therefore interpreted as the ones
corresponding to the disappearance of the barrier. They intersect
in the point on the curve 3, where both barriers are suppressed,
so that the vortex exit and entrance must become nearly
reversible.

In the present paper, we used the method of dimensional estimates,
based on manipulations with two main parameters of the barrier,
namely, barrier width and height. In order to provide an
additional support for our results, we also applied the approach
of Ref.~\cite{Ivlev}, which was initially developed to address
quantum creep in layered high-$T_{c}$ materials. Within this
method, the effective action is first calculated in the
high-temperature regime, which is purely due to TA. Then, the
crossover temperature $T_{0}$ is found by solving the equation of
motion taking into account only the first Fourier harmonic as a
correction to the static solution. For $T < T_{0}$, the action is
estimated, again, by using the TA expression but at $T=T_{0}$. We
found that the results of the method used here turn out to be
quite similar to the results of this approach. For instance,
$T_{0}$ can be obtained from Eq.~(25) of Ref.~\cite{Ivlev} by
dropping the irrelevant elastic term, as $k_{B}T_{0}=-\hbar
U^{\prime\prime}(\rho_{c})/2\pi \eta$. Since
$U^{\prime\prime}(\rho_{c}) \approx -V_{b}/l_{b}^{2}$, we arrive
at the similar result for $T_{0}$. By evaluating
$U^{\prime\prime}(\rho_{c})$, one can also easily show that
$k_{B}T_{0}\sim (\Phi-\Phi_{0})$, in agreement with curve 4 in
Fig.~2. The difference in numerical prefactors of $T_{0}$, as
found by two approaches, is essentially compensated by a very
sharp slope of $T_{0}$ as a function of $\Phi$.

The important question is, how the predicted features can be
tested experimentally. It is possible to detect presence of a
vortex in a disk by using STM methods, as already has been done
in Ref.~\cite{Japanese,Dima} for smaller and thicker disks. In
particular, Ref.~\cite{Japanese} studied the difference between
magnetic fields for which vortex entry and exit occur. It was
shown that this quantity is significantly smaller than it could
be expected from the theory when disregarding fluctuations.
In our point of view, it is attractive to address the temperature
dependence for average fields of vortex entry and exit. One may
assume that the barrier becomes suppressed when the exponent
approaches some critical value, the reasonable estimate being 10.
As seen from Fig. 2, in the absence of fluctuations, the flux of
vortex exit must be independent on $T$, while in the presence of
fluctuations it must grow with increasing $T$. At low $T$, the
corresponding value of magnetic flux must be close to $\Phi
_{out}$. By expanding the r.h.s. of Eqs. (\ref{roc}), (\ref{Uin}),
and (\ref{Uout}) in powers of $\Phi -\Phi _{out}$, one can easily
show that the width of the fluctuative region where thermal
fluctuations are dominant, varies with temperature as $\sim
\sqrt{T}$. The same applies to the case of vortex entry.
Contrarily, when quantum fluctuations are dominant, the width of
the fluctuative region is nearly $T$-independent (except for a
rather weak dependence through $\xi(T)$ and $\lambda(T)$), so that
it tends to a constant at $T=0$. Hence, by analyzing the low-$T$
behavior, one might distinguish between two different mechanisms
of barrier suppression through fluctuations. For the vortex exit,
we predict a crossover between QT and TA (see the upper dotted
curve in Fig.~2). In this case, the slopes of the dependencies of
the effective critical flux on $T$, which are nearly linear, must
be different for low and high $T$.

Perhaps, a more elaborate method to study TA of a \textit{single}
vortex in ultrathin islands has been proposed very recently in
Ref. \cite{Bul1}, where it was concluded that islands of this kind
provide an ideal platform to study such phenomena. It was
suggested to change the applied field periodically and slowly and
to measure not only the average fields for vortex entry/exit, but
also their distributions. The distribution profile must broaden
with the increase of $T$ in the case of TA. This feature provides
an additional tool to distinguish between TA and QT scenarios.

In summary, we have studied the possibility of
%for
both the macroscopic quantum tunnelling and thermal activation of
Abrikosov vortex through the surface barrier in ultrathin
superconducting islands made of Pb and consisting of just few
monoatomic layers. Such islands can be fabricated and studied
using modern techniques. We have found that the barrier for vortex
entry, as well as for the exit, can be suppressed in this system,
even not necessarily in the immediate vicinity of threshold values
of magnetic flux in the absence of fluctuations. The dominant
mechanism of vortex fluctuative entry/exit is determined by the
interplay between the barrier height and width, which are the main
quantities defining probabilities of thermal activation and
quantum tunnelling in a dissipative limit, respectively. A ``phase
diagram'' of the island in the plane of magnetic flux and
temperature was constructed.

This work was supported by the ``Odysseus'' Program of the Flemish Government and the Flemish Science Foundation (FWO-Vl).
W.V.P. acknowledges numerous discussions with A. O. Sboychakov and the support from the Dynasty Foundation, the RFBR (project no. 12-02-00339), and RFBR-CNRS programme (project no. 12-02-91055).

\end{document}